\documentclass[amsmath,nobibnotes,aps,amssymb,pra, prl,aps,showpacs,superscriptaddress,twocolumn, longbibliography, reprint]{revtex4-2}
\usepackage{amsmath,amsfonts,amssymb,amsthm,graphics,graphicx,epsfig,bbm}
\usepackage[colorlinks=true,citecolor=blue,linkcolor=blue,urlcolor=blue]{hyperref}
\usepackage[usenames]{color}
\usepackage{graphicx}
\usepackage{subfigure}
\usepackage{amsmath}
\usepackage{tikz}
\usetikzlibrary{quantikz}
\usepackage{epsfig}
\usepackage{dcolumn}
\usepackage{bm}
\usepackage{color}
\usepackage{times}
\usepackage{epstopdf}
\usepackage{amssymb}
\usepackage{amstext}
\usepackage{latexsym}
\usepackage{float}
\usepackage{hyperref}
\usepackage{amsfonts}
\usepackage{psfrag}
\usepackage{soul,xcolor}
\usepackage[normalem]{ulem}
\usepackage{dsfont}
\usepackage{txfonts}
\usepackage{physics}
\usepackage{footnote}
\usepackage{multirow}
\usepackage{appendix}
\usepackage{mathtools}
\usepackage{xspace}     

\theoremstyle{definition}

\newtheorem*{lemma*}{lemma}
\newtheorem*{corollary*}{Corollary}

\newcommand{\blue}[1]{\textcolor{blue}{#1}}

\usepackage{mathtools}

\begin{document}
\setstcolor{red}
\newtheorem{Proposition}{Proposition}[section]	
%\title{Information Scrambling in a Weakly Kicked Harmonic Oscillator} 
\title{Instabilities in a Non-KAM System via Information Scrambling: A Note} 
\date{\today}

\author{Naga Dileep Varikuti}
\email{dileep.varikuti@unitn.it}
\affiliation{Pitaevskii BEC Center, CNR-INO and Dipartimento di Fisica, Universit\`a di Trento, Via Sommarive 14, Trento, I-38123, Italy}	
\affiliation{INFN-TIFPA, Trento Institute for Fundamental Physics and Applications, Via Sommarive 14, Trento, I-38123, Italy}		
%\affiliation{Department of Physics, Indian Institute of Technology Madras, Chennai, India, 600036}

\begin{abstract}
%\textcolor{red}{We investigate the dynamics of operator growth via out-of-time ordered correlators (OTOCs) in a quantized non-KAM classical system. We take the kicked harmonic oscillator (KHO) as our model. The classical harmonic oscillator system is degenerate when unperturbed. Hence, the KAM theorem remains inapplicable. Resonances are salient features of non-KAM systems. The resonance condition is satisfied whenever the ratio of the frequencies of harmonic oscillator and the kick becomes an integer. Although the Lyapunov exponent at resonance is zero, making the system non-chaotic in the conventional sense, the classical phase space undergoes significant structural changes. We show that the OTOCs are sensitive to these structural changes, which grow quadratically at resonances as opposed to linear growth at non-resonances. We show that the growth of the OTOCs is correlated with the mean energy growth of the system. We give analytical expressions for the OTOCs and show that the time scales of the OTOCs are related to the Euler totient of the frequencies ratio. Our study extends the study of operator growth to the class of non-KAM systems and we discuss possible applications to the quantum metrology.}  

We study operator growth in quantized non-KAM systems using out-of-time-ordered correlators (OTOCs), focusing on the kicked harmonic oscillator as a representative example. Since the classical harmonic oscillator is degenerate, the dynamics fall outside the usual Kolmogorov-Arnold-Moser (KAM) framework, and resonances play a central role in shaping the phase space. We examine the system near resonances, where the ratio between the oscillator and driving frequencies takes integer values. Even though the classical Lyapunov exponent remains small at these points, and hence no conventional chaos, the phase space still undergoes strong structural changes. The OTOCs are particularly sensitive to these resonances, with a quadratic-in-time growth at resonance compared to linear growth away from it. Within a perturbative treatment, we derive closed-form expressions for the OTOCs and uncover a number-theoretic structure emerging in the behavior of OTOCs, governed by the Euler totient function of the frequency ratio. Overall, the results we present in this short note imply that resonant structures can play an important role in controlling information spreading.

\end{abstract}

\maketitle
\section{Introduction}

A classical integrable system with $n$ degrees of freedom evolves on invariant $n$-tori in phase space. The persistence of these tori under weak perturbations is described by the Kolmogorov--Arnold--Moser (KAM) theorem, which guarantees that a large measure of invariant tori survive sufficiently small generic perturbations, up to minute deformations~\cite{poschel2009lecture}. 
%\cite{arnold2009proof, kolmogorov1954conservation, moser1962invariant, moser1967convergent}. 
The applicability of the theorem relies on two crucial assumptions: (i) the integrable Hamiltonian is non-degenerate, and (ii) the motion on the tori satisfies a non-resonance condition, requiring sufficiently irrational frequency ratios. When these conditions fail, the KAM construction no longer holds, and the system can become highly sensitive to even weak perturbations. In particular, resonances corresponding to commensurate frequency ratios lead to abrupt and qualitative changes in phase-space structure, in contrast to the gradual destruction of invariant tori typical of the KAM regime. More generally, systems outside the KAM setting may exhibit strong sensitivity to infinitesimal perturbations and can display chaos-like features even when standard assumptions under which chaos appears are violated \cite{sankaranarayanan2001quantum, sankaranarayanan2001chaos}. 

\iffalse
The OTOCs were first introduced in the context of superconductivity \cite{larkin} and have been recently revived in the literature to study information scrambling in many-body quantum systems \cite{ope2, ope4, lin2018out}, quantum chaos \cite{chaos1, pawan, seshadri2018tripartite, lakshminarayan2019out, shenker2, moudgalya2019operator, manybody2, chaos2, prakash2020scrambling, prakash2019out, markovic2022detecting}, many-body localization \cite{manybody3, manybody4, manybody1, huang2017out} and holographic systems\cite{shock1, shenker3}. For any two operators $A$ and $B$, the out-of-time-ordered commutator function over an arbitrary quantum state $|\psi\rangle$ is given by
\begin{equation}\label{commutator}
C_{AB}(t)=\langle\psi| [A(t), B]^{\dagger}[A(t), B]|\psi\rangle,
\end{equation}
where $A(t)=\hat{U}^{\dagger}(t)A\hat{U}(t)$ is the Heisenberg evolution of $A$ under the Hamiltonian evolution of the system. T 
\fi

In the quantum regime, non-KAM systems may exhibit distinctive signatures such as sharp variations in spectral properties and eigenstate statistics. A paradigmatic example is the kicked harmonic oscillator (KHO), in which the unperturbed harmonic oscillator is degenerate, and the KAM theorem does not apply. As a consequence, invariant tori are not generically robust under perturbations, and resonant conditions, as defined by integer relations between the oscillator and driving frequencies, can play a central role in shaping the dynamics. Even in situations where conventional indicators of chaos, such as the Lyapunov exponent, nearly vanish (for instance, at very weak perturbations), the phase-space structure can still undergo pronounced reorganizations, indicating a strong dynamical sensitivity.

Motivated by these features, we investigate information scrambling in the quantum KHO with emphasis on resonant regimes. The primary diagnostic we employ is the out-of-time-ordered correlator (OTOC), originally introduced in the context of superconductivity \cite{larkin} and now widely used to characterize operator growth and chaos in many-body quantum systems \cite{ope2, lin2018out, shenker2, chaos1, pawan, seshadri2018tripartite, lakshminarayan2019out, prakash2020scrambling, varikuti2022out, PhysRevE.109.014209, wntd-53rd, PhysRevLett.120.040402, PhysRevLett.111.207202}. For two operators $A$ and $B$, we define the commutator-based OTOC as
$
C_{AB}(t)=\langle\psi| [A(t), B]^{\dagger}[A(t), B]|\psi\rangle,
$
where $A(t)=\hat{U}^\dagger(t) A \hat{U}(t)$ denotes Heisenberg evolution under the system dynamics. The function $C_{AB}(t)$ contains two- and four-point correlators. The four-point correlators have an unusual time ordering, hence called OTOCs. In this work, we use the commutator function and OTOCs interchangeably to denote $C_{AB}(t)$. To understand how OTOCs diagnose chaos, we consider position $\hat{X}$ and momentum $\hat{P}$ in the semiclassical limit $(\hbar \to 0)$, where commutators are replaced by Poisson brackets. In this regime, one obtains $\{X(t), P\}^2 \sim \left({\delta X(t)}/{\delta X(0)}\right)^2 \approx e^{2\lambda t}$, where $\lambda$ denotes the Lyapunov exponent (LE), which is positive for classically chaotic dynamics. The correspondence principle then implies that OTOCs in systems with chaotic classical limits exhibit exponential growth up to the Ehrenfest time $t_{EF}$. Beyond $t_{EF}$, quantum corrections dominate, and the semiclassical correspondence breaks down. The OTOCs were shown to be intimately connected to other diagnostics of quantum chaos, including tripartite mutual information \cite{pawan} and the Loschmidt echo \cite{yan2020information}. In addition, they are also related to several quantum information-theoretic resource quantifiers, such as non-stabilizerness and asymmetry~\cite{PhysRevLett.128.050402, varikuti2025deep}, to name a few. In this work, we take the kicked harmonic oscillator (KHO) model as a benchmark model, examine the OTOCs, and demonstrate how the non-KAM features emerge in them.

This note is structured as follows. In Sec.~\ref{section-2}, we review classical and quantum features of the kicked harmonic oscillator model. We present the main results in Sec.~\ref{section-3}, and conclude this work in Sec.~\ref{section-4}. The appendix contains relevant technical details supporting the main text.

\section{Model: Kicked harmonic oscillator}
\label{section-2}

\textit{Classical dynamics.---}
%The KHO model is described by the following Hamiltonian consisting of a particle in the harmonic potential subjected to periodic kicks~\cite{chernikov1989symmetry, billam2009quantum, berman1991problem, kells2005quantum, afanasiev1990width, reichl2021transition, gardiner1997quantum, rechester1980calculation, ichikawa1987stochastic, ishizaki1991anomalous, daly1994classical, borgonovi1995translational, engel2007quantum}:
The KHO model is described by the following Hamiltonian consisting of a particle in the harmonic potential subjected to periodic kicks~\cite{billam2009quantum, kells2005quantum, PhysRevE.109.014209}:
\begin{eqnarray}
H=H_0+ K\cos(kX)\sum_{n=-\infty}^{\infty}\delta\left(t-n\tau\right),
\end{eqnarray}
where $H_0 = ({P^2}/{2m}) + (m\omega^2X^2/2)$ is the Hamiltonian of the unperturbed harmonic oscillator, $m$ is the mass of the particle, $\omega$ denotes the natural frequency of the oscillator, $K$ represents the strength of the kicking potential, $\tau$ is the time interval between successive kicks, and $k$ denotes the wave vector. For simplicity, we fix $m=k=1$ throughout the paper. In the action-angle coordinates, the classical harmonic oscillator system is degenerate, meaning that $H_0$ is linear in the action variable. Hence, $H_0$ is a non-KAM integrable system, and its phase space is not guaranteed to be stable under weak perturbations. This is particularly true if the perturbation is highly non-linear and time-dependent. In the case of the KHO, for certain instances called classical resonances, even arbitrarily small values of $K$ have the potential to induce large-scale structural changes in the dynamics.

\begin{figure}
\includegraphics[scale=0.35]{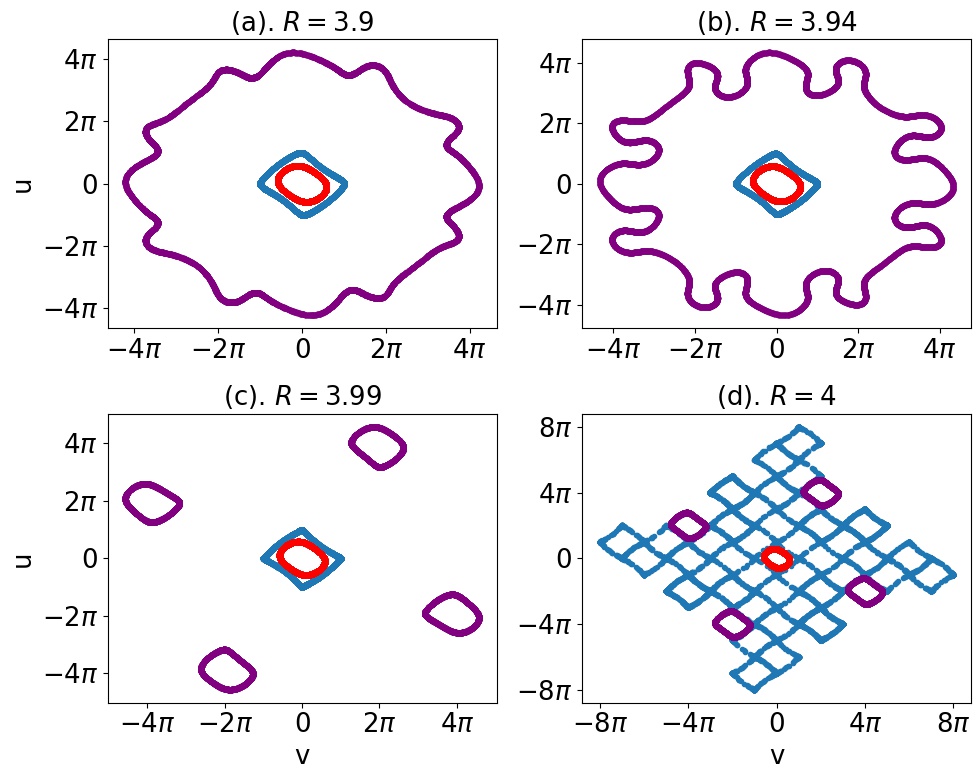}% Here is how to import EPS art
\caption{\label{fig:poincare}The figure depicts representative classical trajectories of the KHO near the resonance condition $R=4$, obtained from three different initial conditions and evolved for $10^4$ time steps. Throughout, the parameters are fixed at $K=1$ and $\tau=1$. At exact resonance, the phase space is organized into three distinct structures: a stochastic web spanning a large portion of phase space, period-four stability islands, and a central period-one regular island surrounding the origin. The chosen initial conditions are selected from each of these regions. As $R$ is varied from $3.9$ toward $4$, following the sequence shown in panels (a)--(d), the trajectories undergo progressive deformation and eventually fragment upon reaching resonance. The stability islands are centered around stable periodic orbits and are enclosed by unstable periodic points. Together, the boundaries of these islands form the stochastic web, which enables transport and diffusion over large distances in phase space.}
\end{figure}

The resonances appear whenever $\omega$ divides $2\pi/\tau$, the kicking frequency, i.e., $\omega R=2\pi/\tau$ with $R\in \mathbb{Z}^{+}$. In other words, the resonances emerge in the system whenever $R$ takes an integer value. For non-integer $R$ values, the system is said to be non-resonant, with the phase space mostly regular and trajectories only slightly deformed. The dynamics of the KHO can be realized using the following two-dimensional dynamical map:
\begin{eqnarray}\label{dynmap}
u_{n+1} &=&(u_n+\epsilon\sin v_n)\cos(\omega\tau) +v_n \sin(\omega\tau),\nonumber\\
v_{n+1} &=&-(u_n+\epsilon\sin v_n)\sin(\omega\tau) +v_n \cos(\omega\tau),
\end{eqnarray}
where $u=P/\omega$, $v=X$ and $\epsilon=K/\omega$. We plot the phase-space trajectories of three randomly chosen initial conditions evolving under the map above. The plots are shown in Fig. \ref{fig:poincare}. The figure shows that the trajectories get increasingly deformed as $R$ approaches an integer. At $R=4$, the phase space contains thin structures of square lattice cells, also known as stochastic webs. Trajectories originating on the boundary can diffuse to infinity over time. Away from the resonance, the phase space is regular with distorted circular trajectories.  

\textit{Quantum dynamics.---}In the quantum domain, the dynamics of the KHO is given by the following Floquet operator:
\begin{equation}\label{floquet}
\hat{U}_\tau=\exp\left\{-\dfrac{2\pi i}{R} \hat{a}^\dagger \hat{a}\right\}\exp\left\{-\dfrac{iK}{\hbar}\cos\hat{X} \right\},
\end{equation}
where $\hat{X}=\sqrt{\hbar/2\omega}\;(\hat{a}+\hat{a}^{\dagger})$. The operators $\hat{a}$ and $\hat{a}^{\dagger}$ represent the bosonic ladder operators for the particle trapped in the harmonic potential. Despite having a seemingly simple Hamiltonian, the quantum KHO displays complex phenomena such as unbounded energy growth, tunneling, translation invariance, Anderson localization, etc. under various conditions \cite{berman1991problem, shepelyansky1992quantum, daly1994classical, kells2005quantum, kells2004dynamical, gardiner1997quantum, carvalho2004web, gardiner2000nonlinear, duffy2004nonlinear, billam2009quantum, borgonovi1995translational, frasca1997quantum}. Furthermore, stochastic webs in the classical phase space can have far-reaching consequences on the corresponding quantum dynamics. %In this work, we examine the operator growth by initializing the system in highly excited Fock states (or number states) of the Harmonic oscillator.

Before turning to the OTOC analysis, it is useful to first examine the evolution of the ladder operators under the quantum KHO dynamics. The Heisenberg evolution of the ladder operators after the first time step can be written as 
\begin{eqnarray}
\hat{a}(1)=\hat{U}^{\dagger}_{\tau}\hat{a}\hat{U}_{\tau}=e^{-2\pi i/R}\left[\hat{a}+\dfrac{iK}{\sqrt{2\hbar\omega}}\sin\hat{X}\right].
\end{eqnarray}
Then, after $t$-number of recursive applications, the time-evolved operator $\hat{a}(t)$ satisfies the following equation:
\begin{eqnarray}\label{bosonic}
\hat{a}(t)e^{2\pi i t/R}=\hat{a}+\dfrac{iK}{\sqrt{2\hbar\omega}}\sum_{j=0}^{t-1}e^{2\pi ij/R}\sin\hat{X}(j),  
\end{eqnarray}
where $\hat{X}(j)=\hat{U}^{\dagger j}_{\tau}\hat{a}\hat{U}^{j}_{\tau}$. Several useful properties of the quantum KHO model, such as mean energy growth and information scrambling, can be deduced by carefully studying the above equation. The phases inside the summation on the right-hand side lead to various operator dynamics depending on whether $R$ assumes an integer or a non-integer value. For instance, if $R=4$, an integer, then the phases are given by $\{\pm 1, \pm i\}$, leading to a coherent summation of the terms $\{\sin\hat{X}(j)$ for all $j\}$. However, an irrational $R$ induces incoherent summations. While the former typically enhances operator growth, the latter suppresses the dynamics, exhibiting behaviors reminiscent of classical diffusion in phase space. In the following section, we carefully examine the OTOCs for the ladder operators.

\iffalse
\begin{figure}
\includegraphics[scale=0.34]{otoc-coarse-fine}% Here is how to import EPS art
\caption{\label{fig:protoc} The OTOCs are sensitive to the changes as the system undergoes a transition from KAM to the non-KAM regime, which is demonstrated for $R=4$ and $R=6$. (a). The plots of OTOC for the parameter values $R=3.9$, $4$, and $4.1$ are drawn. Note that if the decimal point is removed, the numbers $3.9$ and $4.1$ will be transformed to $39$ and $41$, where $41$ is a prime number. (b). Shows the same plot for shorter times. (c). The OTOC are plotted for $R=5.9$ and $R=6$. Corresponding short-time growth is shown in (d). The perturbation strength is kept fixed at $K=0.1$.}
\end{figure}
\fi

\section{Results --- OTOC at Resonance versus Non-resonance}
\label{section-3}

Here, we consider the ladder operators $\hat{a}$ and $\hat{a}^\dagger$ and analyze the corresponding OTOC, focusing primarily on the weak perturbative regime ($K \ll 1$). Owing to the strongly nonlinear nature of the kicking potential $\cos \hat{X}$, an analytical treatment of the OTOCs is, in general, highly difficult. However, simplifications arise in cases where the system displays translational invariance. In particular, for $R \in \{1,2,3,4,6\}$, the OTOCs can be obtained as explicit functions of time by exploiting this symmetry. In the absence of such symmetry, one must rely predominantly on numerical methods to study the dynamics. In Ref.~\cite{PhysRevE.109.014209}, it was shown that at resonances, the OTOCs for the ladder operators typically exhibit quadratic growth over extended time scales, whereas in non-resonant cases the growth is generally suppressed. Here, we present a complementary perspective on operator growth in the same systems. In particular, we show, through a combination of numerical evidence and analytical arguments, that for integer $R$, the OTOC exhibits a piecewise linear growth structure. For a fixed value of $R$, the length $l$ of each linear segment appears to remain constant across segments. Importantly, our results suggest that this segment length is constrained by number-theoretic properties of the system and appears to be bounded from below by Euler’s totient function $\varphi(R)$. On coarse-grained time scales, this piecewise-linear structure gives rise to an overall quadratic envelope for growth. For rational values $R=p/q$, where $p$ and $q$ are coprime integers, the relevant lower bound on the length of the segments is set by $\varphi(p)$. In the following, we shall first numerically examine the OTOCs, averaged over several Fock states, and support the observations with analytical arguments.

\begin{figure}
\includegraphics[scale=0.34]{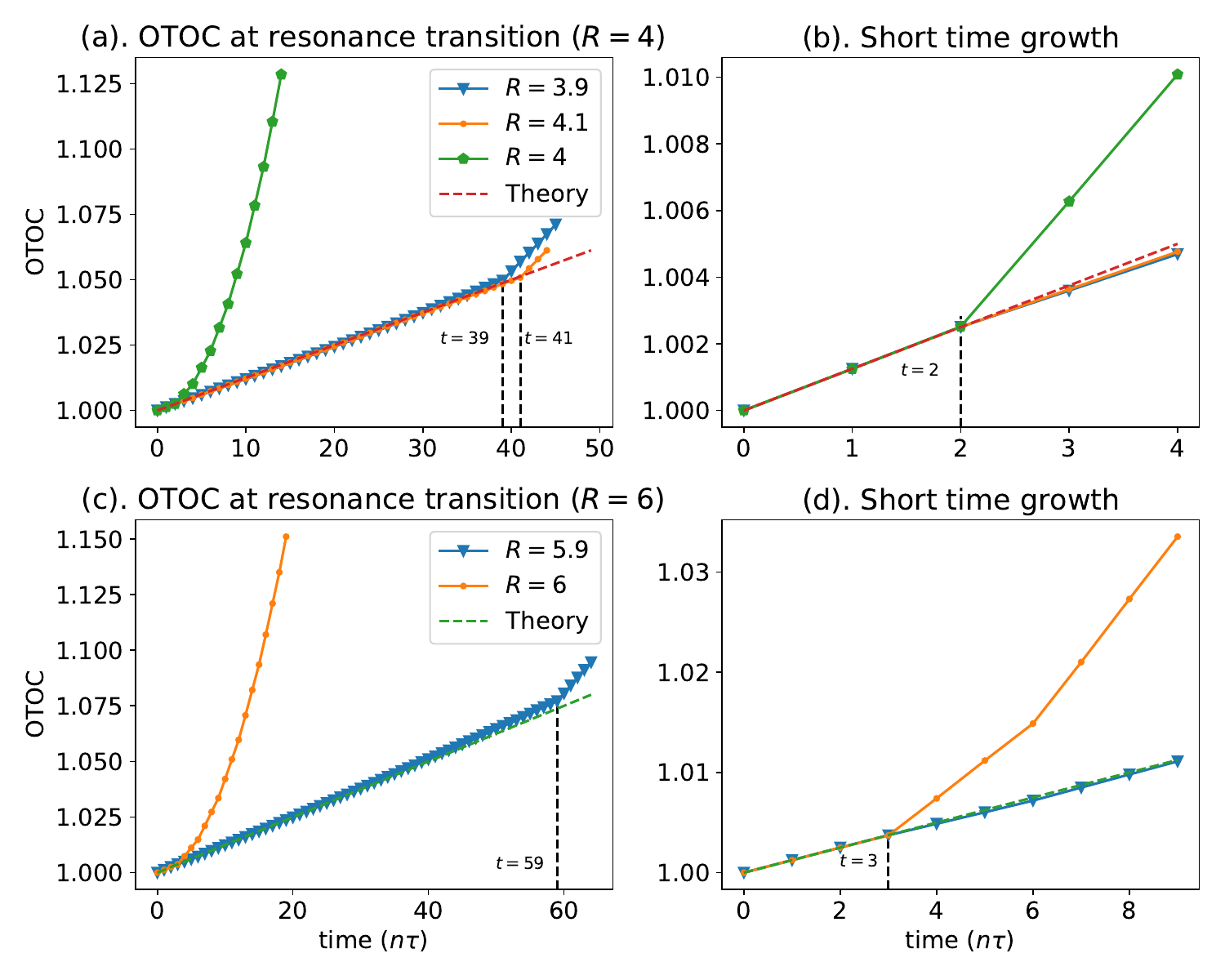}% Here is how to import EPS art
\caption{\label{fig:protoc1} The OTOCs are sensitive to the changes as the system transitions from KAM to the non-KAM regime, which is demonstrated for $R=4$ and $R=6$. (a). The plots of OTOC for the parameter values $R=3.9$, $4$, and $4.1$ are drawn. The values $R=3.9$ and $R=4.1$ correspond to the rational numbers $39/10$ and $41/10$, respectively. For rational $R=p/q$ (with $p$ and $q$ coprime), the relevant arithmetic structure is governed by the numerator $p$. Consequently, the corresponding lower bounds on the linear-growth segment are determined by $\phi(39)$ and $\phi(41)$. The larger value of $\phi(41)$, a consequence of $41$ being prime, accounts for the noticeably extended linear-growth regime observed for $R=4.1$. (b). Shows the same plot for shorter times. (c). The OTOC are plotted for $R=5.9$ and $R=6$. Corresponding short-time growth is shown in (d). The perturbation strength is kept fixed at $K=0.1$.}
\end{figure}

\iffalse
As the numerical results suggest continuous piecewise linear growth for the commutator function $C_{\hat{a}\hat{a}^{\dagger}}(t)$ in a typical coherent state [see Fig. \ref{fig:protoc}b], we here focus mainly on the behavior of the initial linear segment near $t=0$. 
\fi
%Since the excited Fock states exhibit nearly identical behavior over a short period for a fixed $R$. 
We consider the quantity $\overline{C_{\hat{a}\hat{a}^{\dagger}}(t)}$ --- the commutator function averaged over the basis of Fock states. Note here that the averaging involves the first moment of the continuous-variable pure states. Hence, taking the average over the Fock states and the coherent state basis give rise to identical results. We, however, find it convenient to work in the Fock-state basis. Therefore, we are interested in the averaged commutator function given by 
\begin{equation}\label{sf}
\overline{C_{\hat{a}\hat{a}^{\dagger}}(t)}=\lim_{n\rightarrow \infty}\dfrac{1}{n}\sum_{i=0}^{n}\langle n|[\hat{a}(t), \hat{a}^\dagger]^{\dagger}[\hat{a}(t), \hat{a}^\dagger]|n\rangle,  
\end{equation}
For the complete analytical treatment of Eq. (\ref{sf}) in the weak perturbative regime, refer to Appendix \ref{appendix:a}. Here, we briefly mention the steps involved in the derivation. We first assume that $R$ is an integer. Generalization to the non-integers is straightforward. We proceed by writing the following commutator using Eq. (\ref{bosonic}):
\begin{eqnarray}\label{com}
[\hat{a}(t),\thinspace \hat{a}^{\dagger} ] e^{2\pi it/R}=1+\dfrac{iK}{\sqrt{2\hbar \omega}}\sum_{j=0}^{t-1}e^{2\pi ij/R} [\sin\hat{X}(j), \thinspace \hat{a}^{\dagger}]     
\end{eqnarray}
Following this, we write $\sin\hat{X}$ in terms of the displacement operators as $\sin\hat{X}=[D(\alpha)+D^{\dagger}(\alpha)]/2i$, where $\alpha =i/\sqrt{2\omega}$. One can then expand the Heisenberg evolution of the displacement operators under the quantum KHO dynamics in Eq.~(\ref{floquet}) ($\hat{U}^{\dagger j}_{\tau} D(\alpha) \hat{U}^{j}_{\tau}$) using the Taylor series. The resulting expression consists of an infinite series of displacement operators having the linear combinations of roots unity inside their arguments as follows: 
\begin{equation}\label{timeevolvdispmain}
\left[D(\alpha)\right](j)=\sum_{\substack{m_p, n_p=0\\ p=0, ..., j-1}}^{\infty}a_{m_p, n_p}D\left(\alpha\left(\zeta^{j}_{R}+\sum_{p=0}^{j-1}(n_p-m_p) \zeta^{j-1-p}_{R}\right)\right) , 
\end{equation}
where $\zeta_{R}=e^{2\pi i/R}$ denotes the $R$-th root of unity. The coefficients are $a_{m_p, n_p}\equiv a_{m_0, n_0, ..., m_{j-1}, n_{j-1}}$ [see Eq. (\ref{timeevolvdisp}) in the Appendix \ref{appendix:a} for the explicit expressions]. 
In general, given a set of $R$-th roots of unity $\{ \omega^p=e^{2\pi ip/R} \}$, where $p$ covers the integers $\{0, 1, ..., R-1\}$, the largest positive integer $s$ such that $s$ successive roots $1, \omega, \omega^2, ..., \omega^{s-1}$ are linearly independent over the field of rationals $\mathbb{Q}$ is given by the Euler totient of R, $\varphi(R)$. The linear independence implies that 
\begin{eqnarray}            
\sum_{p=0}^{\varphi({R})-1}a_p\omega^p=0\Longleftrightarrow a_p=0\hspace{.1cm}\forall p=0, 1, ..., \varphi(R)-1.
\end{eqnarray}
When the perturbation is very weak, this will lead to the time scale associated with the average commutator function $\overline{C_{\hat{a}\hat{a}^{\dagger}}(t)}$, over which the OTOC grows linearly. This timescale, as determined by the linear independence of the roots, will have a close correspondence with the Euler totient of $R$ --- $\varphi(R)$. To see this, we evaluate the commutator $[\hat{a}(t), \hat{a}^{\dagger}]$ [see Eq. (\ref{com})]. By making use of the identity $[D(\beta), \hat{a}^{\dagger}]=-\beta^* D(\beta)$, $\beta$ is a complex number, we obtain
\begin{widetext}
\begin{equation}\label{commut}
[\hat{a}(t), \hat{a}^{\dagger}]e^{2\pi it/R}=1-\dfrac{iK}{4}\sum_{j=0}^{t-1}e^{2\pi i j/R}\sum_{\substack{m_p, n_p\\ p}}a_{m_p, n_p}\left( e^{-2\pi i j/R}+\sum_{p=0}^{j-1}i_pe^{-2\pi i (j-1-p)/R}\right)\left[D\left(\dfrac{i}{\sqrt{2\omega}}\left( e^{2\pi i j/R}+\sum_{p=0}^{j-1}i_pe^{2\pi i (j-1-p)/R}\right)\right)+\textbf{h.c.}\right].
\end{equation}
\end{widetext}
From Eq. (\ref{commut}), it is straightforward to evaluate the commutator function $|[\hat{a}(t), \hat{a}^{\dagger}]|^2$ followed by the averaging over the Fock state basis. An important step in performing the average is to identify that for a random coherent state $|\alpha\rangle$ and a complex $\beta$, $\overline{\langle\alpha|D(\beta)|\alpha\rangle}=\delta_{\beta, 0}$ --- expectation of the displacement operator averaged over the space of coherent states is non-zero if and only if the argument $\beta$ is zero. In the weak perturbative limit, this linear independence of the roots determines the time scale over which constructive interference persists in the averaged commutator function. Evaluating the commutator explicitly and performing the Fock-state averaging, we obtain that in the limit $K\omega \ll 1$,
\begin{equation}
\overline{C_{\hat{a}\hat{a}^\dagger}(t)} \approx 1 + \frac{K^2 t}{8}, \quad \text{for all } t \leq \varphi(R).
\end{equation}
This result demonstrates that Euler’s totient function $\varphi(R)$ sets a lower bound on the length of the first linear segment $l$ for fixed integer $R$. Moreover, $l$ turns out to be a constant across all the linear segments. The bound $t\leq \varphi(R)$ appears to be sharp for prime and even values of $R$, while for odd composite values, numerical results indicate that $\varphi(R) < l \sim R$. More generally, $t = R$ provides an upper bound on the segment length for all $R \in \mathbb{Z}^+$.

We now compare the numerical results for the OTOCs across different integer values of $R$ with the corresponding theoretical predictions. In Fig.~(\ref{fig:protoc1}), the squared commutator is shown for $R=4$ in the top panel and $R=6$ in the bottom panel. For $R=4$, Euler’s totient function evaluates to $\varphi(4)=2$, and the initial linear growth of the squared commutator persists up to $t=2$. Similarly, for $R=6$, one has $\varphi(6)=2$, while the initial linear regime extends up to $t=3$, confirming that $\varphi(R)$ forms the lower bound for the linear segment of the growth. The cases with non-integer values of $R$ in the vicinity of the respective integer $R$ are also considered in the figure. Remarkably, for $R=3.9$ and $4.1$, the linear segment extends till $t\sim 39$ and $\sim 41$, respectively. This is becuase, for rational frequency ratios $R=p/q$, with $p$ and $q$ coprime, the arithmetic structure governing the initial OTOC growth is determined by the numerator $p$. In particular, the relevant lower bound on the length of the linear-growth segment is set by Euler's totient function $\phi(p)$. Thus, $R=3.9=39/10$ is associated with $\phi(39)=24$, while $R=4.1=41/10$ is associated with $\phi(41)=40$. Since $41$ is prime, $\phi(41)=40$, leading to a substantially longer linear-growth regime compared to the case of $R=4$. This explains why the OTOC behavior can change sharply even for small variations of $R$ near an integer resonance.
%indicating a striking difference between the resonant and non-resonant cases. 
Similarly, in the lower panel, for $R=5.9$, the linear segment extends till $t\sim59 $, confirming the sharp contrast between the two cases.

\section{Conclusion}
\label{section-4}

\iffalse
So far, we have computed the squared commutator functions/OTOCs for the kicked harmonic oscillator analytically and numerically for smaller perturbation strengths where the differences between the resonant and non-resonant cases become very much apparent. We have carefully chosen the creation and annihilation operators to compute the OTOCs. Furthermore, the selection of initial operators also seems to have a profound impact on the behavior of OTOCs. The OTOCs exhibit early linear growth, which lasts for a time that scales with the Euler's Totient function of the integer frequencies ratio. The same has also been observed for the other decimal valued frequency ratios. However, the difference is that in the latter case, the Euler's Totient is a function of the number one obtained after removing the decimal. In the case of irrationals, the linear growth persists for an indefinite amount of time, indicating the suppression of the quantum diffusion.
\fi

In this note, we have computed the OTOCs for the kicked harmonic oscillator both analytically and numerically in the weak perturbation regime, where the distinction between resonant and non-resonant dynamics becomes particularly pronounced. We have specifically chosen the creation and annihilation operators to probe operator growth, and find that the resulting behavior is highly sensitive to whether the system is in resonance or not. In particular, we identified a piecewise linear growth in the OTOCs with quadratic behavior over coarser time scales, with the length of each linear segment governed by Euler’s totient function of the frequency ratio. This analysis highlights a striking s1eparation between resonant and non-resonant driving and demonstrates how number-theoretic properties of the frequency ratio directly impact the scrambling dynamics in the kicked harmonic oscillator.

\iffalse
Looking ahead, similar behavior may arise in many other-body systems, such as collective spin models with generic perturbations, where collective dynamics and finite-size effects can generate analogous structures in operator growth. The observed sensitivity of OTOCs to resonances may also be relevant for quantum simulations, which are typically highly sensitive to imperfections in control, state preparation, and external noise. In such settings, OTOCs provide a useful diagnostic for benchmarking stability and tracking error growth during dynamics. The sensitivity of scrambling dynamics to number-theoretic features of the driving further suggests that even small parameter variations may lead to qualitatively different operator growth. More broadly, understanding whether such arithmetic signatures persist across non-KAM many-body systems could shed light on the interplay between dynamical instability, quantum chaos, and operator spreading.
\fi

It would be interesting to investigate whether similar arithmetic structures emerge in finite dimensional systems such as collective spin models and other many-body systems with generic weak perturbations. The observed sensitivity of OTOCs to resonant structures may also be relevant for quantum simulations of driven systems, where small imperfections in control parameters can significantly affect the dynamics. In such settings, OTOCs may provide a useful diagnostic for benchmarking stability and tracking error growth during dynamics.
Understanding the extent to which such number-theoretic signatures persist in many-body non-KAM systems remains an interesting direction for future work.

\begin{acknowledgments}

This note grew out of earlier work carried out during the author's PhD studies~\cite{PhysRevE.109.014209}, in collaboration with Vaibhav Madhok, Arul Lakshminarayan, and Abinash Sahu. It is a pleasure to acknowledge helpful discussions with Philipp Hauke. This work has benefited from Q@TN, the joint lab between University of Trento, FBK—Fondazione Bruno Kessler, INFN—National Institute for Nuclear Physics, and CNR—National Research Council. The author acknowledges support by Provincia Autonoma di Trento. 
%Views and opinions expressed are, however, those of the author only and do not necessarily reflect those of the European Union or of the Ministry of University and Research.
%Neither the European Union nor the granting authority can be held responsible for them. 

\end{acknowledgments}

\bibliography{main}  

\onecolumngrid
\appendix
\section{Derivation of $[\hat{a}(t), \hat{a}^{\dagger}]$}\label{appendix:a}
The Heisenberg evolution of the bosonic creation and annihilation operators is obtained as
\begin{eqnarray}\label{bosonic-evolution1}
\hat{a}(t)e^{i\omega\tau t}&=&\hat{a}+iK\sqrt{\dfrac{\omega}{2}}\sum_{j=0}^{t-1}e^{ij\omega\tau}\sin\hat{X}(j)\\
&=&\hat{a}+\dfrac{K}{2}\sqrt{\dfrac{\omega}{2}}\sum_{j=0}^{t-1}e^{ij\omega\tau }\left\{U^{\dagger j}\left[D\left(\frac{i}{\sqrt{2\omega}}\right)-D\left(\frac{-i}{\sqrt{2\omega}}\right)\right]U^j\right\}.
\end{eqnarray}
We use Taylor series expansion to obtain Heisenberg evolution of the displacement operators. 
\begin{eqnarray}\label{timeevolvdisp}
U^{\dagger j}D(\alpha)U^j=\sum_{\substack{m_p, n_p=0\\ p=\{0, 1,..., j-1\}}}^{\infty}\left[\prod_{p=0}^{j-1}\dfrac{\left(K\omega \Theta_p\right)^{m_p+n_p}(-1)^{m_p}}{ m_p!n_p!}\right] D\left(\alpha\left(e^{i j\omega\tau}+\sum_{p}(n_p-m_p)e^{i (j-1-p)\omega\tau}\right)\right), 
\end{eqnarray}
where 
\begin{eqnarray}
\Theta_p=\sin\left( \dfrac{\sin((p+1)\omega\tau)}{2\omega\tau}+\sum_{k=0}^{p-1}(n_k-m_k)\dfrac{\sin((k+1)\omega\tau)}{2\omega\tau} \right) 
\end{eqnarray}
We obtain an infinite series expansion for the Heisenberg evolution of the annihilation operator $a(t)$ in terms of the displacement operators as
\begin{eqnarray}
\hat{a}(t)e^{i\omega\tau t}=\hat{a}+\dfrac{K}{2}\sqrt{\dfrac{\omega}{2}}\sum_{j=0}^{t-1}e^{i j\omega\tau}&&\sum_{m_p, n_p, p=\{0, 1, ..., j-1\}}\dfrac{\left(K\omega\Theta_p\right)^{m_p+n_p}(-1)^{m_p}}{ m_p!n_p!}\nonumber\\
&&\left[D\left(\dfrac{i}{\sqrt{2\omega}}\left(e^{i j\omega\tau}+\sum_{p}(n_p-m_p)e^{i (j-1-p)\omega\tau}\right)\right)-D\left(\dfrac{-i}{\sqrt{2\omega}}\left(e^{i j\omega\tau}+\sum_{p}(n_p-m_p)e^{i (j-1-p)\omega\tau}\right)\right)\right].
\end{eqnarray}
The expansion of $\hat{a}(t)$ contains infinite sum of displacement operators. Now, by using the relation $[D(\alpha), \hat{a}^{\dagger}]=-\alpha^*D(\alpha)$, we obtain the commutator function $[\hat{a}(t), \hat{a}^{\dagger}]e^{i\omega\tau t}$ in terms of the displacement operators having linear combinations of $R$-th roots of unity in their arguments. The expression for the commutator is given as follows. 
%Hence, it is possible to compute the commutator $[\hat{a}(t), \hat{a}^\dagger]$. For any complex $\beta$, $[D^\dagger(\beta), a^\dagger]=\beta^* D^\dagger(\beta)$. As a result,
\iffalse
\begin{eqnarray}
\left[D^{\dagger}\left(\pm\alpha\left( e^{i j\omega\tau}+\sum_{k=0}^{j-1}(n_k-m_k)e^{i (j-1-k)\omega\tau}\right)\right), \hat{a}^\dagger\right]=\pm\alpha^{*}\left( e^{-i j\omega\tau}+\sum_{k=0}^{j-1}(n_k-m_k)e^{-i (j-1-k)\omega\tau}\right)D^{\dagger}\left(\pm\alpha\left( e^{i j\omega\tau}+\sum_{k=0}^{j-1}(n_k-m_k)e^{i (j-1-k)\omega\tau}\right)\right).\nonumber\\
\end{eqnarray}
Having equipped with all the essential tools, we now compute the commutator $[a(t), 
a^{\dagger}]$.
\fi
\begin{eqnarray}\label{com1}
[\hat{a}(t), \hat{a}^{\dagger}]e^{i\omega\tau t}=1-\dfrac{iK}{4}\sum_{j=0}^{t-1}e^{i j\omega\tau}&&\sum_{m_p, n_p, p=\{0, 1,..., j-1\}}\dfrac{\left(\omega K\Theta_p\right)^{m_p+n_p}(-1)^{m_p}
}{ m_p!n_p!}\left( e^{-i j\omega\tau}+\sum_{p}(n_p-m_p)e^{-i (j-1-p)\omega\tau}\right)\nonumber\\
&&\left[D\left(\dfrac{i}{\sqrt{2\omega}}\left( e^{i j\omega\tau}+\sum_{p}(n_p-m_p)e^{i (j-1-p)\omega\tau}\right)\right)+D\left(\dfrac{-i}{\sqrt{2\omega}}\left( e^{i j\omega\tau}+\sum_{p}(n_p-m_p)e^{i (j-1-p)\omega\tau}\right)\right)\right].
\end{eqnarray}      
Under certain conditions such as when the perturbation is small, this expression helps us get crucial insights into the nature of initial growth of the squared commutator function. 
%We identify a correspondence between initial growth and the linear independence of $R$-th roots of unity. As a result, we are able to differentiate the KAM regions from non-KAM regions of the kicked harmonic oscillator. An interesting situation appears when $R$ takes the values of irrationals, where the squared commutator functions can be shown to exhibit an indefinite linear growth, which might indicate Anderson type localization in the KHO \cite{kells2005quantum}. 
In what follows, we assume that $t\leq \varphi({R})$ and carry out the calculations. The intution behind this consideration is that the arguments of the displacement operators in the above expression are simply the linear combinations of the roots of unitary of the form $e^{ij\omega\tau}$ and $j$ ranges between $0$ to $t-1$. Therefore, taking $\varphi(R)\leq t$ simplifies the subsequent computations. Then, the squared commutator function averaged over complete set of coherent states becomes
\iffalse 
\begin{eqnarray}
\overline{C_{\hat{a}\hat{a}^{\dagger}}(t)}=1+\dfrac{K^2}{8}\sum_{j=0}^{t-1}\sum_{\substack{m_p, n_p\\ m'_p, n'_p}}\dfrac{\left(K\omega\right)^{m_p+n_p+m'_p+n'_p}(\Theta_p)^{m_p+n_p}(\Theta'_p)^{m'_p+n'_p}(-1)^{m_p+m'_p}}{m_p!n_p!m'_p!n'_p!}\left|e^{-i j\omega\tau}+\sum_{p}(n_p-m_p)e^{-i (j-1-p)\omega\tau}\right|^2\delta_{n_p-m_p, n'_p-m'_p}
\end{eqnarray}
\fi
The above expression transforms as
\begin{eqnarray}
\overline{C_{\hat{a}\hat{a}^{\dagger}}(t)}&=&1+\dfrac{K^2}{8}\sum_{j=0}^{t-1}\sum_{i_p=-\infty}^{+\infty}\left(\sum_{m_p=0}^{+\infty}\dfrac{(K\omega)^{2m_p+i_p}}{m_p!(m_p+i_p)!}(-1)^{m_p} \left( \Theta_{p} \right)^{2m_p+i_p} \right)^2\left| e^{-i j\omega\tau}+\sum_{p}i_pe^{-i (j-1-p)\omega\tau}\right|^2.
\end{eqnarray}
Where we substituted $n_p-m_p=n'_p-m'_p=i_p$ for all $p=\{0, 1, ..., j-1\}$. The summation inside the open bracket adjacent to sum over $p_i$ can be written as a Bessel function of the first kind. 
\begin{eqnarray}
\sum_{m_p=0}^{\infty}\dfrac{(K\omega)^{2m_p+i_p}}{m_p!(m_p+i_p)!}(-1)^{m_p}\left( \Theta_{p} \right)^{2m_p+i_p}=J_{i_p}\left(2 \omega K\Theta_p \right)\nonumber\\
\end{eqnarray}
We now write the squared commutator function using Bessel functions as
\begin{eqnarray}
\overline{C_{\hat{a}\hat{a}^{\dagger}}(t)}=1+\dfrac{K^2}{8}\sum_{j=0}^{t-1}\sum_{i_p=-\infty}^{\infty}J^2_{i_p}\left(2 \omega K \Theta_p \right)\left|e^{-i j\omega\tau}+\sum_{p}i_pe^{-i (j-1-p)\omega\tau}\right|^2,
\end{eqnarray}
At $j^{th}$ time-step the averaged mod squared commutator function grows by the quantity 
\begin{eqnarray}\label{otoattj}
C(t=j)&=&\dfrac{K^2}{8}\sum_{i_p=-\infty}^{\infty}J^2_{i_p}\left(2 \omega K\Theta_p \right)\left|e^{-i j\omega\tau}+\sum_{p}i_pe^{-i (j-1-p)\omega\tau}\right|^2 \nonumber\\
&=&\dfrac{K^2}{8}+O(K^4).
\end{eqnarray}
While approximating the Eq. (\ref{otoattj}) we made use of the following Bessel function identities 
\begin{eqnarray}
\sum_{p_i=-\infty}^{\infty}J_{p_i}^2(\beta)=1\hspace{0.5cm}\text{and} \hspace{0.5cm} \sum_{p_i=-\infty}^{\infty}p_i^2J_{p_i}(\beta)=\dfrac{\beta^2}{2}.
\end{eqnarray}
Since the perturbation strength $K$ is small, the higher order terms in $K$ can be ignored and we finally obtain
\begin{eqnarray}
\overline{C_{\hat{a}\hat{a}^{\dagger}}(t)}=1+\dfrac{K^2}{8}t \hspace{1cm}(t\leq \phi(R)), 
\end{eqnarray}
where $\phi(R)$ is the Euler totient- which computes the number of coprimes less than $R$.

\end{document}